\documentclass[aps,prb,twocolumn,showpacs]{revtex4-1}  
\usepackage{graphicx,graphics,color,epsfig} 
\usepackage{amsmath,amssymb,amsfonts} 
\usepackage{bm,xspace,mhchem,times} 
\usepackage[colorlinks,citecolor=blue,linkcolor=red]{hyperref} 

\begin{document}

%\title{What is Phonon Spin and Elastic Spin? A Revisit}
\title{From Elastic Spin to Phonon Spin: Symmetry and Fundamental Relations}
%\title{Elastic Spin and Orbital Angular Momentum, Phonon Spin and Orbital Angular Momentum: Symmetry and Fundamental Relations}
%\title{From Elastic Spin to Phonon Spin: Topological Material and Lattice Examples}
\author{Jie Ren}
\email{Corresponding Email: Xonics@tongji.edu.cn}
\affiliation{%
Center for Phononics and Thermal Energy Science, China-EU Joint Lab on Nanophononics, Shanghai Key Laboratory of Special Artificial Microstructure Materials and Technology, School of Physics Science and Engineering, Tongji University, Shanghai 200092, China
}%

\date{\today}
\pacs{63.20.-e, 62.30.+d, 47.10.ab, 11.30.-j, 03.70.+k}

\begin{abstract}
This note is mainly based on postgraduate lectures at Tongji University since 2020 spring. 
We firstly revisit the elastic spin and orbital angular momentum [PNAS 115, 9951 (2018)] but more general for anisotropic systems by applying the Noether's theorem to the elastic Lagrangian, and by applying the symmetry argument in the field theory. Then, fundamental relations between elastic energy flux and elastic spin are uncovered. In particular cases, the wave spin is closely related to the vorticity of energy flux and momentum. Secondly, we move forward from the elastic spin to revisit the phonon spin [Fizika Tverdogo Tela 3, 2160 (1961)] by applying the second quantization to elastic fields. We show that the uncovered phonon spin, a polarized elastic-vibration quanta, is generally not restricted to transverse phonon modes, but applying to general phonon modes, like the longitudinal phonon modes, surface phonon modes, hybridized phonon modes and so on, regarded as a consequence of mode interferences. The elastic spin and phonon spin originate from the local rotating of the field polarization in time domain, not the local circulation (vorticity) of displacement or velocity in space domain. Hope the results could advance the fundamental understanding of phonon spin and elastic spin, and help people to promote the spin phononics for hybrid quantum sensing and technology with multiple degrees of freedom.
\end{abstract}

\maketitle
{%\color{blue}
%({\bf Introduction}
%Introducing the background, the history. What the scientists have achieved right now. Where does the problem comes from? What is the significance of the question? Why it is important.)

Since the brilliant insight of P. A. M. Dirac on the relativistic quantum theory in the year 1928, the study of spin and orbital angular momentum had been extended to arbitrary fields by Belinfante around 1940~\cite{Belinfante1940}. Then the field theory became a powerful tool to study the energy, momentum, and angular momentum of various (scalar, vector, spinor) fields described by different Lagrangians. It was gradually accepted as a common knowledge~\cite{soper2008classical} that a vector field can possess intrinsic spin in additional to orbital angular momentum when applying Noether's variational method upon rotational invariance~\cite{Noether1918}. 
When studying phonons, the quanta of vibration field in solids, the inherent spin was being overlooked. It was until 1961, Vonsovskii and Svirskii firstly reported in Ref.~\cite{PhononSpin} (written in Russian with an English translation in 1962 in Ref.~\cite{PhononSpin1}) that the spin of transverse phonon in isotropic medium can be expected and formulated, which inspired Levine's study concerning quantized spin of transverse phonons in isotropic cubic lattice in 1962~\cite{Levine1962}. These pioneering works were followed by recent revived discussions of the angular momentum physics of circular polarized phonons~\cite{CPphonon, Zhang2014PRL, NakanePRB2018}, as well as in the standard textbook~\cite{Auld}, where the Faraday rotation and rotary activity of shear waves were discussed. 

However, the initial works Refs.~\cite{PhononSpin,PhononSpin1,Levine1962} only uncovered the phonon spin of transverse (shear) modes, taking the analogy of photon spin of circularly polarized light~\cite{Auld}, but concluded the phonons of longitudinal modes have zero spin. 
This was also regarded to be consistent with the classical field theory because the longitudinal curl-free field can be usually regarded as a scalar field since the gradient of a scalar field is always irrotational without vorticity, and the classical field theory predicted that the scalar field has no intrinsic spin~\cite{soper2008classical}. Therefore, it was always thought that the wave spin is impossible in the longitudinal curl-free field, such as the acoustic wave in fluid (air, water).  

It is until 2018, half a century later, people start to realize that~\cite{Ren2018arXiv1, Ren2018arXiv2, long2018intrinsic} the Helmholtz decomposition theorem, {\it any well-behaved vector field can be decomposed into the sum of a transverse (non-diverging, solenoidal) vector field and a longitudinal (non-curling, irrotational) vector field}, enables us to produce versatile interference patterns among different components that are spin free individually to synthesize rich wave spin texture. It is therein recognized that the longitudinal (elastic, acoustic) wave can have nonzero spin~\cite{Ren2018arXiv1,Ren2018arXiv2}. Not only that, it shows mixed transverse-longitudinal waves can induce the hybrid spin, which is responsible for rich wave spin phenomena beyond pure transverse waves~\cite{long2018intrinsic}, such as the later experimental observation of spin-momentum locking in surface acoustic waves excited with chiral elastic meta-sources~\cite{yuan2021observation}.  

Since then, the spin angular momentum of the elastic wave is termed as the {\it elastic spin} for short, and attracts much attention~\cite{ValleyanisotropyPRB2019,Geilen2020APL,Brun2020IJES} as well as the elastic orbital angular momentum counterpart~\cite{ChaplainPRL2022,Bliokharxiv2022,Deymier_2018}. Moreover, the multi-physical interactions between elastic spin of surface acoustic wave and the underlying magnetic materials and piezoelectric materials are attracting more and more attention~\cite{Zhang2020PRL,Zhao2020PRApplied,Yu2021PRB, Sasaki2021NC,Sonner2021SA,Sensors2021}. 
The {\it acoustic spin}~\cite{Ren2018arXiv2} is the pure longitudinal wave version of elastic spin and phonon spin, %elastic spin~\cite{long2018intrinsic} and phonon spin~\cite{PhononSpin,PhononSpin1,Levine1962}, 
proposed in Refs.~\cite{Ren2018arXiv1, Ren2018arXiv2}, firstly observed in experiments~\cite{shi2019observation}, and triggered an upsurge of acoustic spin in complicated structural acoustics~\cite{bliokh2019spin,bliokh2019transverse,Rondon_2020}, acoustic spin torque~\cite{Toftul2019PRL, Yang2021JAP} and the acoustic spin induced selective directionality and scattering with potential applications~\cite{Long2020NSR, Wei2020NJP, Long2020NC}.

%~\cite{bliokh2019spin}
%~\cite{bliokh2019transverse}

%~\cite{Toftul2019PRL} % acoustic spin torque
%~\cite{Yang2021JAP} % acoustic spin torque in TIs.

%~\cite{Rondon_2020} %Acoustic vortex beams in synthetic magnetic fields
%~\cite{Wei2020NJP} % Far-field and near-field directionality in acoustic scattering
%
%~\cite{Long2020NC}
%~\cite{Long2020NSR}

%~\cite{ChaplainPRL2022} %Elastic Orbital Angular Momentum. with an Erratum.
%~\cite{Bliokharxiv2022}

%~\cite{Zhang2020PRL} % Unidirectional Pumping of Phonons by Magnetization Dynamics
%~\cite{Zhao2020PRApplied} %Phonon Transport Controlled by Ferromagnetic Resonance
%~\cite{Yu2021PRB} %Dynamic magnetoelastic boundary conditions and the pumping of phonons
%~\cite{Sasaki2021NC} %Magnetization control by angular momentum transfer from surface acoustic wave to ferromagnetic spin moments
%~\cite{Sonner2021SA} %Ultrafast electron cycloids driven by the transverse spin of a surface acoustic wave
%~\cite{Sensors2021} %The Experimental Registration of the Evanescent Acoustic Wave in YX LiNbO3 Plate

%~\cite{ValleyanisotropyPRB2019} %Valley anisotropy in elastic metamaterials
%~\cite{Geilen2020APL} %Interference of co-propagating Rayleigh and Sezawa waves observed with micro-focused Brillouin light scattering spectroscopy
%~\cite{Brun2020IJES} %Rayleigh waves in micro-structured elastic systems: Non-reciprocity and energy symmetry breaking

Here, we plan to revisit the elastic spin and phonon spin from a consistent perspective and exhibit their fundamental relations. This note is structured as followed: In Section \ref{SecI} we revisit the elastic spin and orbital angular momentum from the symmetry and conservation argument in the field theory, by applying the Noether's theorem to the Lagrangian of the elastic system. It is followed by detailed discussions about symmetry and properties of elastic spin and orbital angular momentum in Section \ref{SecII}. We then discuss intrinsic relations between elastic energy flux and elastic spin in Section \ref{SecIII}, showing that not only will the vorticity of elastic spin have influence on the elastic energy flux, but in turn the vorticity of energy flux and momentum will have a close connection to the elastic spin.
Starting from Section  \ref{SecIV}, we move forward to discuss general phonon spin by applying the second quantization to elastic fields. The phonon spin picture uncovered here is not restricted to the transverse phonon modes, but applying to general phonon modes, like the longitudinal phonon modes, surface phonon modes and hybridized phonon modes, whose symmetries and properties are discussed in Section \ref{SecV}. We finally conclude in Section \ref{SecVI} with some brief discussions and perspectives. 

\section{Elastic Spin and Orbital Angular Momentum}\label{SecI}
Elastic spin and orbital angular momentum (AM) are previously discussed in isotropic medium. Here, we revisit the problem in a general sense beyond the isotropic constraint, by applying Noether's theorem~\cite{Noether1918} to the infinitesimal invariance.
Considering the elastic wave propagating in a general anisotropic solid, the local deformation at $x^\mu$ in the solid is a vector field $u_i(x^\mu)$. The equation of motion of this displacement field reads:
\begin{equation}
\label{EOM}
\rho \ddot u_i = \frac{\partial }{\partial x^j}C_{ijkl}\frac{\partial  u_k}{\partial x^l},
\end{equation}
where $\rho$ is the density and $C_{ijkl}$ is the Christoffel matrix of the material. The Lagrangian (density) description of the elastic system is~\cite{Landau,laude2015lagrangian}:
\begin{equation}
\label{eq_lag}
\mathcal{L}(u_i,\partial_\mu u_i, x^\mu) = \frac 12 \left( \rho \dot  u_i^2 - C_{ijkl} \frac{\partial u_{i}}{\partial x^j} \frac{\partial u_{k}}{\partial x^l} \right), 
\end{equation}
%which is a typical Lagrangian for vector field with up to first order derivatives. 
whose Euler-Lagrangian equation $\frac{\partial \mathcal{L}}{\partial u_i}=\partial_\mu \frac{\partial \mathcal{L}}{\partial (\partial_\mu u_i)}$ leads to the elastic equation Eq.~\eqref{EOM}. 
Using Noether's theorem, we are able to obtain the energy-momentum tensor and AM possessed by this field, following the standard procedure~\cite{soper2008classical}.

Firstly, we consider an infinitesimal spacetime translation by $\delta x^\mu=a^\mu$ such that $x^\mu \rightarrow x^\mu + a^\mu$, which induces the field variation: $\delta u_i =  - \frac{\partial u_i}{\partial x^\mu} \delta x^\mu=-(\partial_{\mu} u_i)a^\mu$. As such, the variation of Lagrangian can be written as:
\begin{equation}
\delta \mathcal{L} = -\partial_\mu \left(\frac{\partial \mathcal{L}}{\partial (\partial_\mu u_i)} \frac{\partial u_i}{\partial x^\nu} - \delta_{\nu}^{\mu} \mathcal{L}  \right) a^\nu.
\end{equation}
Therefore, the invariance of Lagrangian $\delta \mathcal{L}=0$ upon arbitrary infinitesimal spacetime translation $a^\nu$ leads to the conservation $\partial_\mu T^{\nu\mu}=0$, where $T^{\nu\mu}$ is so-called energy-momentum tensor:
\begin{equation}
T^{\nu\mu} = \frac{\partial \mathcal{L}}{\partial (\partial_\mu u_i)} {\partial^\nu u_i} - g^{\nu\mu} \mathcal{L},
\end{equation}
with $g^{\nu\mu}$ being the flat spacetime metric ${\rm diag} (1,-1,-1,-1)$. For elastic system, this energy-momentum tensor, as the conserved Noether current associated with spacetime translations, gives the total energy (Hamiltonian) density:
\begin{equation}
H=T^{00}=  \frac{\partial \mathcal{L}}{\partial (\partial_0 u_i)} {\partial_0 u_i} - \mathcal{L}=\frac 12 \left( \rho \dot  u_i^2 + C_{ijkl} \frac{\partial u_{i}}{\partial x^j} \frac{\partial u_{k}}{\partial x^l} \right),
\end{equation}
and the $k$-direction linear (orbital) momentum density:  
\begin{equation}
p^l_k=T^{k0}=-\frac{\partial \mathcal{L}}{\partial (\partial_0 u_i)} {\partial_k u_i}=-\rho \dot  u_i \partial_k u_i.
\end{equation}
Note that this general expression in anisotropic elastic medium keeps as the same as the expression in specific isotropic medium in Refs.~\cite{PhononSpin,PhononSpin1}. 
The linear momentum density can be written as a more compact vector form: 
\begin{equation}
\bm{p}^l=-\rho \bm{\dot u}\cdot (\nabla) \bm{u}, 
\end{equation}
which on period average is equivalent to $\bm {p}^l=\frac{\rho\omega}{2} \text{Re}[\bm{u}^*\cdot (-i\nabla) \bm{u}]=\frac{\rho\omega}{2} \text{Im}[\bm{u}^*\cdot (\nabla) \bm{u}]$ in frequency domain, as the same as obtained in Ref.~\cite{long2018intrinsic}. 

Moreover, the divergenceless property of the tensor $\partial_\mu T^{0\mu}=0$ leads to the energy continuity relation: 
\begin{equation}
\frac{\partial H}{\partial t}+\nabla\cdot \bm P=0,
\end{equation}
i.e., $\frac{\partial T^{00}}{\partial t}+\partial_j P_j=0$ with the energy flux  $P_j=T^{0j}=\frac{\partial \mathcal{L}}{\partial (\partial_j u_i)}\dot{u}_i=-C_{ijkl}\frac{\partial u_k}{\partial x^l}\dot u_i$ being just the $j$-direction component of the Poynting vector $\bm P=-\bm\sigma\cdot\bm{\dot u}$.  Here $\bm \sigma=\mathbf C: \bm \varepsilon$ is the Cauchy stress tensor, as the product of stiffness tensor $\mathbf C$ and strain tensor $\bm \varepsilon$. The divergenceless property of the tensor $\partial_\mu T^{k\mu}=0$ leads to the momentum continuity relation: 
\begin{equation}
\frac{\partial \bm p^l}{\partial t}+\nabla\cdot \bm T=0,
\end{equation}
i.e., $\frac{\partial T^{k0}}{\partial t}+\partial_j T^{kj}=0$, so that the stress tensor $\bm T$ with elements $T^{kj}=-\frac{\partial \mathcal{L}}{\partial (\partial_j u_i)} {\partial_k u_i}+\delta_{kj}\mathcal{L}$ has the clear physical meaning of stress (shear stress when $k\neq j$). 
It is worth noting that the Poynting vector $\bm P$ as energy flux has a different dimension from the linear momentum density $\bm {p}^l$, with an additional factor of the square of velocity. %Therefore, when using the orbital component of Poynting energy flux ($\bm r\times\bm P^o$) to calculate the elastic OAM density, the energy flux needs to be normalized by the square of velocity~\cite{long2018intrinsic} $[\text{velocity}]^2={C_{ijkl}}/{\rho}$ since $[\text{position}]\times[\text{Poynting energy flux}]=[\text{position}]\times[\text{momentum density}]\cdot[\text{velocity}]^2$.

Then, let us consider the space rotation of the elastic system by an infinitesimal angle $\bm \theta$ with coordinate change: $x^k\rightarrow x^k+\theta_i x^j \epsilon_{ijk}$. In this case, the field varies not only with respect to the change of coordinates $\delta x^k=\theta_i x^j \epsilon_{ijk}$, but also due to the orientation change of the field itself, so that the total variation of the field is recorded as:
\begin{equation}
\delta u_{j}=-\frac{\partial u_{j}}{\partial x^k}\delta x^k+\theta_i u_k\epsilon_{ikj}. % =-\theta_i (\epsilon_{ijk}x^j\frac{\partial u_{j}}{\partial x^k}+\epsilon_{ijk}u_k).
\end{equation}
This expression tells us an important fact that the rotation of the elastic field, not only changes the field distribution upon coordinate change (the first term that will lead to OAM), but also changes the orientation of the field polarization itself (the second term that will lead to SAM).  
As such, the variation of Lagrangian $\delta \mathcal{L}=\partial_{\mu}(\frac{\partial \mathcal{L}}{\partial(\partial_{\mu}u_j)}\delta u_j)+\partial_{\mu}\mathcal{L}\delta x^{\mu}$ under spatial rotation can be derived as:
\begin{eqnarray}\label{totalAM}
\delta \mathcal{L} &=& \theta_i \partial_\mu J_{i\mu}, \;\;\;\; \text{where:}  \nonumber \\
J_{i\mu}&=& \epsilon_{ijk} x^j T^{k\mu }+ \epsilon_{ikj} u_k\frac{\partial \mathcal{L}}{\partial(\partial_\mu u_j )}.
\end{eqnarray}
The invariance $\delta\mathcal L=0$ upon arbitrary infinitesimal rotation $\theta_i$ leads to the conserved Noether current $J_{i0}$, which is the total AM. %corresponding to the rotational symmetry. 
Clearly, the total AM separates into two parts: $J_{i0}=L_i+S_i$, where the first part
\begin{equation}\label{OAM_component}
L_i =  \epsilon_{ijk} x^j T^{k0} = - \epsilon_{ijk}x^j \rho\dot{u}_m \partial_k u_m
\end{equation}
is the elastic OAM with the {\it extrinsic} dependence $\bm r=\{x^1,x^2,x^3\}$ on the choice of reference origin. While the second part 
\begin{equation}\label{SAM_component}
S_i=\epsilon_{ikj} u_k\frac{\partial \mathcal{L}}{\partial(\partial_0 u_j )}=\epsilon_{ikj}\rho u_k \dot{u}_j
\end{equation}
is the elastic SAM as a local property of the field itself, {\it intrinsic} in the sense that it is independent on the reference origin and thus named spin AM. 

Note that, the elastic spin describes a local rotating of the field polarization in time domain, rather than the vorticity of the displacement field in space domain, so that the elastic spin $\bm S$ has nothing to do with the spatial curl $\nabla \times \bm u$. Indeed, for the longitudinal wave $\bm u=\nabla \Phi$ described by a gradient of a scalar field $\Phi$, the vorticity is absent $(\nabla \times \nabla \Phi=0)$ with no doubt. Yet, the longitudinal wave can still possess nontrivial wave spin $\bm S$~\cite{Ren2018arXiv1, Ren2018arXiv2, long2018intrinsic, shi2019observation}. Moreover, different from previous studies that are restricted to isotropic homogeneous medium, the derivation of elastic spin and OAM here breaks the limitation and applies to more general anisotropic inhomogeneous cases. For instance, we can see that the anisotropic medium considered here has the same expression of elastic spin as the isotropic medium reported in Refs.~\cite{PhononSpin,PhononSpin1}. This is also reflected by the fact that the final expressions of local elastic spin density and elastic OAM density do not contain the information of the anisotropic elasticity tensor, but with only dependence on the local field $\bm u(\bm r), \dot {\bm u}(\bm r)$ and local density $\rho(\bm r)$.  Anisotropy may not conserve the OAM and SAM, but the definitions are still there.  
%Finally we note that, $L_i$ and $S_i$ may not be conserved individually, e.g., when the energy-momentum tensor is not symmetric, but the total AM $J_{i0}$ can still be well conserved in total.

It is worth noting that the Poynting vector $\bm P$ as energy flux has a factor of velocity square $[\text{velocity}]^2={C_{ijkl}}/{\rho}$ in addition to the linear momentum density $\bm {p}^l$. Therefore, when using the orbital component of Poynting energy flux ($\bm r\times\bm P^o$) to calculate the elastic OAM density ($\bm r\times\bm p^l$), the energy flux needs to be normalized by the square of velocity $[\text{velocity}]^2={C_{ijkl}}/{\rho}$ since the OAM is $[\text{position}]\times[\text{momentum density}]=[\text{position}]\times[\text{Poynting energy flux}]/[\text{velocity}]^2$. The elastic OAM obtained in the main text of Ref.~\cite{long2018intrinsic} is also discussed in Ref.~\cite{ChaplainPRL2022} and its erratum, as well as Ref.~\cite{Bliokharxiv2022}. %In fact, the Eq. (12) in Ref.~\cite{ChaplainPRL2022} does not need Erratum since it clearly states that it is the ``energy flux density'', more precisely, the orbital component of $\bm P^o$, ~\cite{long2018intrinsic}

\section{Symmetry and Properties of Elastic Spin and Elastic OAM}\label{SecII}
 
The elastic OAM Eq.~\eqref{OAM_component} and SAM density Eq.~\eqref{SAM_component} can be rewritten in a more compact vector form:
\begin{eqnarray}
\bm L&=&\rho \bm{\dot u}\cdot (-\bm r\times\nabla) \bm{u} =\bm r\times\bm{p}^l, \\
\bm S&=&\rho \bm{u}\times  \bm{\dot u},
\end{eqnarray}
which on period average for monochromatic elastic wave $\bm u \sim e^{-i\omega t}$ are equivalent to 
\begin{eqnarray}
\bm L&=&\frac{\rho\omega}{2} \bm r\times\text{Im}[\bm{u}^*\cdot (\nabla) \bm{u}], \\
\bm S&=&\frac{\rho\omega}{2} \text{Im}[\bm{u}^*\times\bm{u}].
\end{eqnarray}
These results exactly reproduce the same expressions as obtained in Ref.~\cite{long2018intrinsic}, where general waves were considered with exemplifications in isotropic elasticity, and are here further extended to general solids with anisotropic properties.  
Also, it was pointed out that~\cite{long2018intrinsic} one can define state vectors: 
\begin{equation}
\begin{aligned}
|\bm{u}\rangle = \sqrt{\frac{\rho\omega}{2\hbar}}\begin{pmatrix}
u_{x} \\
u_{y} \\
u_{z} 
\end{pmatrix}, \quad
\langle\bm{u}|= \sqrt{\frac{\rho\omega}{2\hbar}}\begin{pmatrix}
u_{x}^* & u_{y}^* & u_{z}^*
\end{pmatrix}.
\end{aligned}
\end{equation}
As such, $\langle\bm{u}|\bm{u}\rangle=\frac{\rho \omega^2\sum_i|u_i|^2}{2}/(\hbar\omega)$ has the physical meaning of phonon number density, 
and with the momentum operator $\hat{\bm p}=-i\hbar\nabla$, the previous discussed linear momentum can be apparently expressed as $\bm p^l={\rm Re}[\langle\bm{u}|\hat{\bm{p}}|\bm{u}\rangle]$. 
Therefore, the period-averaged elastic spin and orbital AM density can be accordingly expressed as ``local expectation''-like values in convenient quantum-like representation, as: 
\begin{eqnarray}
\bm S(\bm r)&=&\langle\bm{u}(\bm r)|\hat{\bm{S}}|\bm{u}(\bm r)\rangle,  \\
\bm L(\bm r)&=&{\rm Re}[\langle\bm{u}(\bm r)|\hat{\bm{L}}|\bm{u}(\bm r)\rangle],
\end{eqnarray}
where $\hat{\bm{L}}=\bm{r}\times(-i\hbar\nabla)$ can be regarded as the OAM operator, satisfying commutation $[\hat{L}_i,\hat{L}_j]=i \hbar\epsilon_{ijk} \hat{L}_k$, while $\hat{\bm{S}}$ is the spin-1 operator, 
\begin{equation}\label{Spin1}
\begin{aligned}
\hat{S}_{x} &= -i \hbar
\begin{pmatrix}
0 & 0 & 0\\
0 &  0 & 1\\
0 & -1 & 0
\end{pmatrix}, \quad
\hat{S}_{y} &= -i \hbar
\begin{pmatrix}
0 & 0 & -1\\
0 &  0 & 0\\
1 & 0 & 0
\end{pmatrix}, \\
\hat{S}_{z} &= -i \hbar
\begin{pmatrix}
0 & 1 & 0\\
-1 &  0 & 0\\
0 & 0 & 0
\end{pmatrix}
\end{aligned}
\end{equation}
satisfying $[\hat{S}_i,\hat{S}_j]=i \hbar\epsilon_{ijk} \hat{S}_k$. This spin-1 representation is consistent with the fact that the elastic vibration field can be quantized into a bosonic field, so called phonon.
% so that both elastic spin and phonon spin can be regarded as the consequence of  invariance in $SO(3)$ group. 

Strictly speaking, Noether theorem and symmetry discussions in the previous section can be conventionally imposed within the Lorentz group in four dimension spacetime~\cite{soper2008classical, gravitation}, so that elastic spin will emerge as a consequence of the Lorentz invariance. This is a standard technique in applying Noether's theorem to discussing wave spin in other classical waves, such as stratified fluid wave~\cite{Jones1973, WaterSpin2022}, plasma wave~\cite{Jones1973}, and gravitational wave~\cite{gravitation, evanscent, Xin2021}. In above, without loss of generality, we restricted discussions to the rotation in three dimension space for simplicity, so that elastic spin emerges as the consequence of the $SO(3)$ symmetry.   
Indeed, expressing the infinitesimal $\bm \theta$-rotation as $\hat R=\left(\begin{array}{ccc}1 & -\theta_z & \theta_y \\\theta_z & 1 & -\theta_x \\-\theta_y & \theta_x & 1\end{array}\right)$ by keeping only upto the first order $\bm \theta$, one can see that the $SO(3)$ transformation $\hat U_{\bm J}$ of the elastic field not only rotates the spatial coordinates %$\hat R \bm r=\bm r+\bm\theta\times \bm r$ 
but also rotates the elastic field states, as:
\begin{eqnarray}
&&\hat U_{\bm J}|\bm u(\bm r)\rangle=\hat R |\bm u(\hat R^{-1}\bm r)\rangle  \nonumber \\
&=& |\bm u(\bm r-\bm\theta\times\bm r)\rangle+\bm \theta\times|\bm u(\bm r-\bm\theta\times\bm r)\rangle   \nonumber \\
&\approx&|\bm u(\bm r)\rangle-(\bm\theta\times\bm r)\cdot \nabla|\bm u(\bm r)\rangle+\bm \theta\times|\bm u(\bm r)\rangle  \nonumber\\
&=&|\bm u(\bm r)\rangle-\bm\theta\cdot(\bm r\times \nabla)|\bm u(\bm r)\rangle+\bm \theta\times|\bm u(\bm r)\rangle   \nonumber\\
&=&\big (1-\frac{i}{\hbar}\bm\theta\cdot (\hat{\bm{L}}+ \hat{\bm{S}})\big )|\bm u(\bm r)\rangle \approx e^{-\frac{i}{\hbar}\bm\theta\cdot \hat{\bm{J}}}|\bm u(\bm r)\rangle.
%&\approx&e^{-\frac{i}{\hbar}\bm\theta\cdot (\hat{\bm{L}}+ \hat{\bm{S}}))}|\bm u(\bm r)\rangle
\end{eqnarray}
This follows the standard technique in quantum mechanics textbook~\cite{Messiah}, leading to the infinitesimal $SO(3)$ transformation
$\hat U_{\bm J}\approx e^{-\frac{i}{\hbar}\bm\theta\cdot \hat{\bm{J}}}$, where the infinitesimal generator $\hat{\bm{J}}=\hat{\bm{L}}+ \hat{\bm{S}}$ can be regarded as the total AM operator of the elastic field. From this argument, one can also clearly see that the elastic OAM ${\bm{L}}$ results from redistributing the field upon coordinate change, while the elastic SAM ${\bm{S}}$ results from the orientation change of the field polarization (displacement or velocity) without coordinate change.  

{%\color{red}
For general monochromatic elastic wave at $\bm r=\{x,y,z\}$, $\bm u(\bm r)=\{|u_x|e^{i\phi_x},|u_y|e^{i\phi_y},|u_z|e^{i\phi_z}\}e^{-i\omega t}$, the linear momentum density is related to the phase gradient of the elastic field, as
\begin{equation}\label{eq:phaseG}
\bm {p}^l=\frac{\rho\omega}{2} \sum_{i=x,y,z}e^{-2\text{Im}[\phi_i]}|u_i|^2\text{Re}[\nabla \phi_i].
\end{equation}
Using Eq.~(\ref{eq:phaseG}), we can see that the elastic OAM is related to the phase gradient of the displacement field 
\begin{equation}
\bm L=\frac{\rho\omega}{2} \sum_{i=x,y,z}e^{-2\text{Im}[\phi_i]}|u_i|^2 (\bm r\times\nabla) \text{Re}[\phi_i]. 
\end{equation}
For example, for elastic field in 2D membrane system with polar coordinate and real phase $\phi_i$, if only the in-plane modes are excited, then 
\begin{equation}
L_z=\frac{\rho\omega}{2}(|u_x|^2 \partial_\theta \phi_x+|u_y|^2 \partial_\theta \phi_y), 
\end{equation}
while if only the out-of-plane modes are excited, then 
\begin{equation}
L_z=\frac{\rho\omega}{2}|u_z|^2 \partial_\theta \phi_z,
\end{equation}
where the phase gradient $\partial_{\theta}\phi$ can be quantized as an integer OAM number $l$ ($\phi\sim l\theta$), with $l$ so-called topological charge.
{%\color{blue} 
The phase gradient expression indicates that the elastic OAM $\bm L$ is related to the physical momentum around the origin, since the phase gradient is related to the linear momentum density, which is bounded by the group velocity of the wave in the medium.}  

%Distinctly, the elastic spin describes a local rotating of the field polarization in time domain, rather than the vorticity of the field in space domain, so that the elastic spin $\bm S$ has nothing to do with the curl $\nabla \times \bm u$. Indeed, for the longitudinal wave $\bm u=\nabla \Phi$ described by a gradient of a scalar field $\Phi$, the vorticity is absent $\nabla \times \bm u=0$ without no doubt, yet, the longitudinal wave can still possess nontrivial wave spin $\bm S$~\cite{long2018intrinsic}. 
Accordingly, for general elastic wave $\bm u(\bm r)=\{|u_x|e^{i\phi_x},|u_y|e^{i\phi_y},|u_z|e^{i\phi_z}\}e^{-i\omega t}$, the elastic spin $\bm S(\bm r)=\{S_x, S_y, S_z\}$ can be obtained as
\begin{subequations}
\begin{align}
S_x(\bm r)&=\rho \omega|u_y(\bm r)u_z(\bm r)|\cos(\phi_y(\bm r)-\phi_z(\bm r)), \\
S_y(\bm r)&=\rho \omega|u_z(\bm r)u_x(\bm r)|\cos(\phi_z(\bm r)-\phi_x(\bm r)),  \\
S_z(\bm r)&=\rho \omega|u_x(\bm r)u_y(\bm r)|\cos(\phi_x(\bm r)-\phi_y(\bm r)),
\end{align}
\end{subequations}
which can be also regarded as the consequence of wave interference~\cite{long2018intrinsic}, manifested as the phase difference from different polarization components at location $\bm r$~\cite{Xin2021}. 
%\color{blue} 
Therefore, distinct from $\bm L$, the polarization rotation, as the physics origin of intrinsic elastic spin $\bm S$, is related to the phase change, so that the polarization rotation speed is related to the phase velocity, but not restricted by the group velocity in the medium.
}

It is interesting to note that here we consider neutral elastic medium without electric charges. If the elastic medium is uniformly charged, the elastic OAM and SAM will induce nonzero elastic orbital magnetic moment and elastic spin magnetic moment by simply multiplying an additional factor of $\frac{\rho_e}{2\rho}$ with $\rho_e$ being the charge density. As such, the elastic orbital magnetic moment density $\bm \mu_L=\frac{\rho_e}{2\rho}\bm L=\frac{\rho_e}{2} \bm{\dot u}\cdot (-\bm r\times\nabla) \bm{u}$ and elastic spin magnetic moment density $\bm \mu_S=\frac{\rho_e}{2\rho}\bm S=\frac{\rho_e}{2}\bm{u}\times  \bm{\dot u}$, which on period average for monochromatic elastic wave $\bm u \sim e^{-i\omega t}$ are equivalent to 
\begin{equation}
\bm \mu_L=\frac{\rho_e\omega}{4} \bm r\times\text{Im}[\bm{u}^*\cdot (\nabla) \bm{u}], \quad \bm \mu_S=\frac{\rho_e\omega}{4} \text{Im}[\bm{u}^*\times\bm{u}].
\end{equation}
Actually, even for the neutral elastic medium, once the positive and negative charges are separate, i.e., polarized, the system can also possess nontrivial elastic magnetic moment.  
The detailed discussion of elastic magnetic moment is out of the scope of this note and will be investigated elsewhere. 

\section{Relations between Elastic Spin and Energy Flux}\label{SecIII}
We expect that there exist intrinsic relations between elastic spin and elastic energy flux because intuitively the polarization rotation may relate to the circulation of energy flux. As we will see in the following, this is true for some simple cases, but is not generally hold.

For simplicity, we restrict discussions of this section in isotropic elastic medium. The stress tensor is expressed as $\bm\sigma=\lambda \nabla\cdot \bm u \bm 1+2\mu\bm\varepsilon$, where the first Lame constant $\lambda=K-\frac{2}{3}G$ and the second Lame constant $\mu=G$. $K$ and $G$ are the bulk and shear modulus, respectively. As such, the equation of motion of the elastic field can be much simplified as:
\begin{equation}
-\rho \omega^2 \bm u=\nabla\cdot \bm \sigma=(\lambda+2\mu) \nabla\nabla\cdot \bm u-\mu \nabla\times\nabla\times \bm u.
\end{equation}
Accordingly, after some mathematics, we can express the period-averaged Poynting energy flux $\bm P=\frac{1}{2}\text{Re}[i\omega \bm \sigma^* \bm u]$ in different ways, as:
\begin{eqnarray}\label{PoyntingRelation}
\bm P&=&\frac{\omega (\lambda+2\mu)}{2}\text{Im}[\bm u^*(\nabla\cdot\bm u)]+\frac{\omega \mu}{2}\text{Im}[\bm u^*\times(\nabla\times \bm u)] \nonumber \\
&& -\frac{\omega\mu}{2}\nabla\times\text{Im}[\bm u^*\times \bm u] \nonumber \\
&=&v^2_L\frac{\rho\omega}{2}\text{Im}[\bm u^*(\nabla\cdot\bm u)]+v^2_T\frac{\rho\omega}{2}\text{Im}[\bm u^*\times(\nabla\times \bm u)]  \nonumber \\
&&-v^2_T\nabla\times \bm S, 
%=&&(v^2_L-v^2_T)\frac{\rho\omega}{2}\text{Im}[\bm u^*(\nabla\cdot\bm u)]+v^2_T (\bm p^l-\bm p^s), 
\end{eqnarray}
where $v_L=\sqrt{(\lambda+2\mu)/\rho}$ and $v_T=\sqrt{\mu/\rho}$ are the longitudinal and transverse wave speeds, respectively. Clearly, the vorticity of elastic spin $\nabla\times\bm S$ will have influence on the elastic energy flux $\bm P$. Here we used the identity $\text{Im}[\bm u^*(\nabla\cdot\bm u)]+\text{Im}[\bm u^*\times(\nabla\times \bm u)]=\text{Im}[\bm u^*\cdot(\nabla)\bm u)]+\frac{1}{2}\nabla\times\text{Im}[\bm u^*\times \bm u]$, which implies an interesting momentum decomposition of complex field with both transverse and longitudinal wave components: $\frac{\rho\omega}{2}\text{Im}[\bm u^*(\nabla\cdot\bm u)]+\frac{\rho\omega}{2}\text{Im}[\bm u^*\times(\nabla\times \bm u)]=\bm p^l+\bm p^s$. $\bm p^l=\frac{\rho\omega}{2} \text{Im}[\bm{u}^*\cdot (\nabla) \bm{u}]$ is the linear momentum density discussed above and the momentum density $\bm p^s=\frac{1}{2}\nabla\times\bm S$ can be named as the spin momentum density. It is worth noting that the volume integration of the cross product between $\bm r$ and the whole momentum density leads to the total AM, as $\int \bm r\times\frac{\rho\omega}{2}(\text{Im}[\bm u^*(\nabla\cdot\bm u)]+\text{Im}[\bm u^*\times(\nabla\times \bm u)])d\bm r^3 =\int \bm r\times(\bm p^l+\bm p^s)d\bm r^3 =\int (\bm L+ \bm S)d \bm r^3$.

To see the potential connection between elastic spin and energy flux,  let us discuss the particular cases of transverse and longitudinal waves separately. Note the notation of transverse (longitudinal) wave we used here does not mean that the wave has displacement only perpendicular (parallel) to its propagation direction, but may have field component along (perpendicular) to the propagation direction, as in the cases of evanescent surface waves or waves with inhomogeneous field distribution. Nevertheless, the transverse (longitudinal) waves are still divergence-free (curl-free). Therefore, the notation of transverse (longitudinal) wave means the divergence-free (curl-free) field throughout the whole work. 

Using the Helmholtz field decomposition $\bm u=\bm u_T+\bm u_L$ into curl-free field $\nabla\times \bm u_L=0$ and divergence-free field $\nabla\cdot \bm u_T=0$, we can obtain the relations between Poynting energy flux and elastic spin and momentum for longitudinal and transverse waves, respectively.  As we will see, not only will the vorticity of elastic spin (spin momentum) have influence on the elastic energy flux, but in turn the vorticity of energy flux and momentum will have a close connection to the elastic spin. For the physical quantity composed of pure transverse (longitudinal) wave field, ``$T(L)$'' is added as the subscript of the notation.
For elastic wave of pure transverse wave component, the elastic energy flux can be decomposed as:
\begin{equation}\label{PT}
%\bm P=\bm P_T&&=v^2_T\frac{\rho\omega}{2}\text{Im}[\bm u_T^*\times(\nabla\times \bm u_T)]-v^2_T\nabla\times \bm S_T  \nonumber\\
%&&= v^2_T(\bm p^l_T+\bm p^s_T)-2 v^2_T\bm p^s_T   \nonumber \\
%&&= v^2_T\bm p^l_T-v^2_T\bm p^s_T.
\bm P_T= v^2_T\bm p^l-\frac{v^2_T}{2}\nabla\times \bm S=v^2_T\bm (\bm p^l-\bm p^s).
\end{equation}
For elastic wave of pure longitudinal wave component, the elastic energy flux can be decomposed as:
\begin{equation}\label{PL}
%\bm P=\bm P_L&&=v^2_L\frac{\rho\omega}{2}\text{Im}[\bm u_L^*(\nabla\cdot\bm u_L)]-v^2_T\nabla\times \bm S_L  \nonumber\\
%&&= v^2_L(\bm p^l_L+\bm p^s_L)-2 v^2_T\bm p^s_L   \nonumber \\
%&&= v^2_L\bm p^l_L+(v^2_L-v^2_T)\bm p^s_L.
\bm P_L= v^2_L(\bm p^l+\frac{1}{2}\nabla\times \bm S)-v^2_T\nabla\times \bm S= v^2_L\bm p^l+(v^2_L-2v^2_T)\bm p^s.
\end{equation}
Moreover, for longitudinal elastic wave, from the relation $\nabla\times[(\nabla\cdot \bm u_L)\bm u^*_L]=\nabla^2\bm u_L\times \bm u^*_L$ and the equation of motion $-\rho\omega^2 \bm u_L=(\lambda+2\mu)\nabla^2 \bm u_L$, it is straightforward to obtain that 
\begin{equation}
\nabla\times\text{Im}[(\nabla\cdot \bm u_L)\bm u^*_L]=\frac{\rho\omega^2}{\lambda+2\mu}\text{Im}[\bm u^*_L\times \bm u_L],
\end{equation}
and the vorticity of linear momentum is related to the spin, as
\begin{equation}
\nabla\times  \bm p^l=\frac{\omega^2}{v_L^2}\bm S+\frac{1}{2}\nabla^2\bm S.
\end{equation}
Therefore, the vorticity of elastic energy flux (Poynting vector) is related to the elastic spin, as:
\begin{equation}
\nabla\times \bm P_L=\omega^2 \bm S+v_T^2 \nabla^2 \bm S.
\end{equation}

For the sound wave in shearless fluid, like air or water, $\mu=0$ and $v_T=0$, so that the Poynting vector $\bm P_L=v_L^2 (\bm p^l+\bm p^s)$ and the acoustic spin is related to the vorticity of Poynting vector, as: 
\begin{equation}
\bm S=\frac{1}{\omega^2}\nabla\times \bm P_L,
\end{equation}
i.e. a local (intrinsic) circulation of energy flux. (Note: to obtain this connection, we used the property $\nabla\cdot \bm S=0$ and $\nabla\times \bm p^s=-\frac{1}{2}\nabla^2 \bm S$ for curl-free longitudinal waves). In other words, the wave spin emerges even for irrotational field $\nabla\times \bm u_L=0$, because the energy flux can have microscopic spatial circulation locally ($\nabla\times \bm P_L\neq 0$) to produce nonzero spin angular momentum at a point. Therefore, it is reasonable to use the vorticity of Poynting energy flux to represent the pseudospin in acoustic systems, and this kind of acoustic pseudospin (energy flux vorticity) is tightly related to the acoustic spin, i.e., the real, physical spin angular momentum of acoustic wave~\cite{shi2019observation}, with a factor difference of the frequency square. 

Nevertheless, this simple connection merely applies to longitudinal (irrotational, shearless, curl-free) waves, but is not strictly applicable to general transverse waves, let alone general cases with hybrid transverse-longitudinal waves. For example, for the pure transverse elastic wave $\nabla\cdot \bm u_T=0$, one needs further restrict to a special elastic field, like the in-plane vibrations that $u_z=0$ and $u_{x,y}$ is $z$-independent, to guarantee $\nabla\cdot\bm S=0$. As such, it is readily to get $\nabla\times\text{Im}[\bm u^*_T\times(\nabla\times\bm u_T)]=\frac{\rho\omega^2}{\mu}\text{Im}[\bm u^*_T\times\bm u_T]$ so that the vorticity of linear momentum is related to the spin as $\nabla \times \bm p^l=\frac{\omega^2}{v^2_T}\bm S+\frac{1}{2}\nabla^2\bm S$. Therefore, the vorticity of energy flux of this special transverse wave can be related to the elastic spin as well, as:
\begin{equation}
\nabla\times \bm P_T=\omega^2 \bm S+v_T^2 \nabla^2 \bm S.
\end{equation}
This is, however, not valid for general transverse waves. 

\section{Phonon Spin by Quantization}\label{SecIV}
Now, we turn to discuss the phonon spin~\cite{PhononSpin,PhononSpin1}, by applying the second quantization of the elastic field to the elastic spin (the second quantization of the elastic OAM to phonon OAM will be discussed elsewhere). At low temperature, or with the low density of phonons, the vibrations of the medium should be described by the laws of quantum mechanics. At high temperature, or the phonon density is sufficiently large, the phonon picture of the elastic system will reduce to the equivalent classical description. 

Through the Boson operators of second quantization, we can express the displacement and velocity vector, as:
\begin{eqnarray}
\bm u&=&\sum_{\bm k\lambda}\sqrt{\frac{\hbar}{2\rho V \omega_{\bm k\lambda}}}(b^{\dag}_{-\bm k\lambda}+b_{\bm k\lambda})\bm\varepsilon_{\bm k\lambda}e^{i\bm k\cdot\bm r}, \\
\dot{\bm u}&=&i\sum_{\bm k\lambda}\sqrt{\frac{\hbar\omega_{\bm k\lambda}}{2\rho V}}(b^{\dag}_{-\bm k\lambda}-b_{\bm k\lambda})\bm\varepsilon_{\bm k\lambda}e^{i\bm k\cdot\bm r},
\end{eqnarray}
where $\bm k$ is the wave vector, $\lambda$ is the branch index (LA, TA, LO, TO and so on), $\omega_{\bm k\lambda}$ is the phonon frequency, $\rho$ is the medium density, and $V$ is the normalized volume.
$\bm\varepsilon_{\bm k\lambda}$ is the orthonormal unit polarization vector and $b^{\dag}_{-\bm k\lambda}, b_{\bm k\lambda}$ are the creation and annihilation operators of corresponding phonons, satisfying the commutation relation $[b_{\bm k\lambda}, b^{\dag}_{\bm k'\lambda'}]=\delta_{\bm k\bm k'}\delta_{\lambda\lambda'}$. Considering $\omega_{-\bm k\lambda}=\omega_{\bm k\lambda}^*$ and $\bm\varepsilon_{-\bm k\lambda}=\bm\varepsilon_{\bm k\lambda}^*$, the linear momentum density $\bm p^l$ can be readily quantized as:
\begin{equation}
\bm p^l=\frac{1}{V}\sum_{\bm k\lambda}\hbar\bm k(b^{\dag}_{\bm k\lambda}b_{\bm k\lambda}+\frac{1}{2}),
\end{equation}
by removing the high frequency terms upon time-average on period, where $\bm k$ can be generally equal to the phase gradient $\nabla \phi$ for the field $\bm u\sim e^{i(\phi-\omega t)}$. Quantization of the elastic field energy is also straightforward.

The elastic spin $\bm S=\rho \bm u\times\dot{\bm u}$ can be expressed in terms of phonon operators as well, as:
\begin{eqnarray}
\bm S=\frac{i\hbar}{2V}&&\sum_{\bm k\bm k'}\sum_{\lambda\lambda'}\sqrt{\frac{\omega_{\bm k'\lambda'}^*}{\omega_{\bm k\lambda}^*}}(\bm\varepsilon^*_{\bm k\lambda}\times\bm\varepsilon^*_{\bm k'\lambda'})e^{-i(\bm k+\bm k')\bm r}b^{\dag}_{\bm k\lambda}b^{\dag}_{\bm k'\lambda'} \nonumber \\
-&&\sqrt{\frac{\omega_{\bm k'\lambda'}}{\omega_{\bm k\lambda}}}(\bm\varepsilon_{\bm k\lambda}\times\bm\varepsilon_{\bm k'\lambda'})e^{i(\bm k+\bm k')\bm r}b_{\bm k\lambda}b_{\bm k'\lambda'} \nonumber \\
+&&\sqrt{\frac{\omega_{\bm k'\lambda'}^*}{\omega_{\bm k\lambda}}}(\bm\varepsilon_{\bm k\lambda}\times\bm\varepsilon^*_{\bm k'\lambda'})e^{i(\bm k-\bm k')\bm r}b_{\bm k\lambda}b^{\dag}_{\bm k'\lambda'} \nonumber \\
-&&\sqrt{\frac{\omega_{\bm k'\lambda'}}{\omega_{\bm k\lambda}^*}}(\bm\varepsilon^*_{\bm k\lambda}\times\bm\varepsilon_{\bm k'\lambda'})e^{-i(\bm k-\bm k')\bm r}b^{\dag}_{\bm k\lambda}b_{\bm k'\lambda'}.
\end{eqnarray}
The contribution of first two terms $b^{\dag}_{\bm k\lambda}b^{\dag}_{\bm k'\lambda'}, b_{\bm k\lambda}b_{\bm k'\lambda'}$ will vanish on the period average, reminiscent to removing the high frequency terms in rotating wave approximation. Moreover, considering the elastic field quantization is on the monochromatic wave $\bm u \sim e^{-i\omega t}$, the square root of the frequency ratios become $1$. As such, the phonon spin is simplified to
\begin{eqnarray}
\bm S=\frac{i\hbar}{2V}\sum_{\bm k\bm k'}\sum_{\lambda\lambda'}&&(\bm\varepsilon_{\bm k\lambda}\times\bm\varepsilon^*_{\bm k'\lambda'})e^{i(\bm k-\bm k')\bm r}b_{\bm k\lambda}b^{\dag}_{\bm k'\lambda'} \nonumber \\
-&&(\bm\varepsilon^*_{\bm k\lambda}\times\bm\varepsilon_{\bm k'\lambda'})e^{-i(\bm k-\bm k')\bm r}b^{\dag}_{\bm k\lambda}b_{\bm k'\lambda'}.
\end{eqnarray}
By considering the spin-1 operator $\hat{\bm{S}}$ in Eq.~\eqref{Spin1} to denote $-i\hbar\bm\varepsilon^*_{\bm k\lambda}\times\bm\varepsilon_{\bm k'\lambda'}=\langle\bm\varepsilon_{\bm k\lambda}|\hat{\bm{S}}|\bm\varepsilon_{\bm k'\lambda'}\rangle$, and exchanging indices $\bm k\leftrightarrow\bm k'$ and $\lambda\leftrightarrow\lambda'$, the phonon spin is finally expressed as:
\begin{equation}
\bm S=\frac{1}{V}\sum_{\bm k\bm k'}\sum_{\lambda\lambda'}e^{-i(\bm k-\bm k')\bm r} \langle\bm\varepsilon_{\bm k\lambda}|\hat{\bm{S}}|\bm\varepsilon_{\bm k'\lambda'}\rangle (b^{\dag}_{\bm k\lambda}b_{\bm k'\lambda'}+\frac{1}{2}\delta_{\bm k\bm k'}\delta_{\lambda\lambda'}).
\end{equation}

Clearly, this expression indicates highly rich physics of phonon spin for diverse cases of phonon hybridizations between different modes $(\bm k, \lambda)$ and $(\bm k', \lambda')$,  beyond the phonon spin of only transverse phonons discussed in previous literatures~\cite{PhononSpin,PhononSpin1,Levine1962}. Just similar to the elastic spin, the phonon spin can also result from the interference~\cite{long2018intrinsic} between longitudinal phonons with different momentum (direction) $\bm k$, or from the interference between longitudinal and transverse phonons (different branch $\lambda$) with the same $\bm k$. For the phonon mode with the unique $\bm k'=\bm k, \lambda'=\lambda$, the zero-point phonon spin will appear but will flip to opposite when $\bm k$ is reversed. For complex unit cell with more than one atom, the system can have complicated phonon bands, like various acoustic and optical branches, so that the situations become more elaborate. In the next section, we will discuss case by case the hybridization and symmetry for phonon spin in general. Specific example of phonon spin in topological materials and lattices will be discussed in a subsequent article. %From Elastic Spin to Phonon Spin: Topological Material and Lattice Examples

\section{Hybridization and Symmetry for Phonon Spin}\label{SecV}

Different phonon modes $(\bm k, \lambda)$ and $(\bm k', \lambda')$ can hybridize to form a new phonon mode, so that even the original phonons individually have zero spin, the phonon spin of the hybrid modes may emerge to be nonzero. 

Firstly, let us set $\bm\varepsilon_{\bm k\lambda}^*\times\bm\varepsilon_{\bm k'\lambda'}=|\bm\varepsilon_{\bm k\lambda}^*\times\bm\varepsilon_{\bm k'\lambda'}|e^{i\varphi}$, so that $\bm\varepsilon_{\bm k\lambda}\times\bm\varepsilon_{\bm k'\lambda'}^*=|\bm\varepsilon_{\bm k\lambda}^*\times\bm\varepsilon_{\bm k'\lambda'}|e^{-i\varphi}$. We denote a new phase $e^{i\phi}=e^{i(\varphi-(\bm k-\bm k')\bm r)}$, and then for a given pair of phonon modes $\langle(\bm k,\lambda), (\bm k',\lambda')\rangle$, the phonon spin is expressed as:
\begin{equation}
\bm S=-i\hbar|\bm\varepsilon_{\bm k\lambda}^*\times\bm\varepsilon_{\bm k'\lambda'}|(e^{i\phi}b^{\dag}_{\bm k\lambda}b_{\bm k'\lambda'}-e^{-i\phi}b_{\bm k\lambda}b^{\dag}_{\bm k'\lambda'}).
\end{equation}
Secondly, let us define new phonons as the hybridization of the paired phonon modes $\langle(\bm k,\lambda), (\bm k',\lambda')\rangle$, through:
\begin{eqnarray}\label{hybridization}
b_{\pm}=\frac{b_{\bm k\lambda}\mp i e^{i\phi}b_{\bm k'\lambda'}}{\sqrt 2}, \quad b^{\dag}_{\pm}=\frac{b^{\dag}_{\bm k\lambda}\pm i e^{-i\phi}b^{\dag}_{\bm k'\lambda'}}{\sqrt 2},
\end{eqnarray}
which, as can be easily checked, follow the same Boson commutators $[b_{\pm},b_{\pm}^{\dag}]=1,[b_{\pm},b_{\mp}^{\dag}]=0$. And it is readily to verify $b_{+}^{\dag}b_{+}-b_{-}^{\dag}b_{-}=-i(e^{i\phi}b^{\dag}_{\bm k\lambda}b_{\bm k'\lambda'}-e^{-i\phi}b_{\bm k\lambda}b^{\dag}_{\bm k'\lambda'})$. Therefore, $b_{+}^{\dag}b_{+}, b_{-}^{\dag}b_{-}$ denote the spinful phonon modes with opposite SAMs, and we have the phonon spin for the paired modes $\langle(\bm k,\lambda), (\bm k',\lambda')\rangle$ (that may be non-spinful individually), as:
\begin{eqnarray}
\bm S&=&\hbar|\bm\varepsilon^*_{\bm k\lambda}\times\bm\varepsilon_{\bm k'\lambda'}|(b_{+}^{\dag}b_{+}-b_{-}^{\dag}b_{-}) \nonumber \\
&=&|\langle\bm\varepsilon_{\bm k\lambda}|\hat{\bm{S}}|\bm\varepsilon_{\bm k'\lambda'}\rangle|(b_{+}^{\dag}b_{+}-b_{-}^{\dag}b_{-}).
\end{eqnarray}
Generally speaking, the phase $\phi$ and hybridization may be spatial direction dependent, and $\bm S$ will be location dependent in the real space, exhibiting the interference pattern as a consequence of the difference between $\bm k$ and $\bm k'$. Since the hybridization can be between arbitrary pair of phonon modes, the polarization is generally elliptical and the value $|\langle\bm\varepsilon_{\bm k\lambda}|\hat{\bm{S}}|\bm\varepsilon_{\bm k'\lambda'}\rangle|=\hbar|\bm\varepsilon^*_{\bm k\lambda}\times\bm\varepsilon_{\bm k'\lambda'}|$ is upper bounded by spin $1$ $(\hbar)$. In general, $\bm S$ will be spatially varying and has rich structure of phonon (elastic) spin texture in both real and momentum space. 

Beside the elliptically polarized hybrid phonons in general, we now discuss some simple cases for the paired phonons with the same $\bm k$ and the degenerate (yet different) branches $\lambda,\lambda'$. For example, for the two degenerate bulk transverse phonons at long wave limit (TA modes), the polarization vectors are real with $\bm\varepsilon_{\bm k\lambda}\times\bm\varepsilon_{\bm k\lambda'}=\frac{\bm k}{k}$ and $\phi=0$. As such, the hybridization Eq.~\eqref{hybridization} reduces to
\begin{eqnarray}
b_{\bm k\pm}=\frac{b_{\bm k\lambda}\mp i b_{\bm k\lambda'}}{\sqrt 2}, \quad b^{\dag}_{\bm k\pm}=\frac{b^{\dag}_{\bm k\lambda}\pm i b^{\dag}_{\bm k\lambda'}}{\sqrt 2}.
\end{eqnarray}
The two linear polarized phonons can be hybridized into two oppositely circular polarized phonons $b_{+}^{\dag}b_{+}, b_{-}^{\dag}b_{-}$, so that the phonon spin reads: 
\begin{eqnarray}
\bm S_{\bm k}=\hbar\frac{\bm k}{k}(b_{\bm k+}^{\dag}b_{\bm k+}-b_{\bm k-}^{\dag}b_{\bm k-}).
\end{eqnarray}
The direction of this phonon spin is parallel to the direction of momentum $\bm k$, which exhibits the phonon helicity and the tight phonon spin-momentum locking that reversing $\bm k$ reverses $\bm S_{\bm k}$. 

Two phonons merely degenerate at particular $\bm k$ points can also have similar nonzero hybrid spin. For example, the 2D Kekule lattice of $C_6$ point group symmetry with in-plane vibrations has two-fold degenerate phonons at $\bm k=0$, i.e., two $p$-orbital modes and two $d$-orbital modes.  Hybridization between linear polarized modes $d_{x^2-y^2}$ and $d_{xy}$ with zero phonon spin leads to the circularly polarized modes $d_{\pm}=(d_{x^2-y^2}\pm i d_{xy})/\sqrt 2$ with nonzero phonon spin:
\begin{eqnarray}
S_{z}=\hbar(b_{d_+}^{\dag}b_{d_+}-b_{d_-}^{\dag}b_{d_-}).
\end{eqnarray}
The case for $p$-orbitals with $p_{\pm}=(p_x\pm ip_y)/\sqrt 2$ is somewhat similar. Also, the circular polarized phonons of hybrid LO-TO modes in graphene~\cite{CPphonon} falls into this phonon spin scenario. In general, when the energies of two phonon branches coincide, the degeneracy points in $\bm k$-space may either be isolated points, or form a line, loop or even plane, which can be regarded as the consequence of crystal lattice symmetry. 

Moreover, the single phonon mode can also possess nonzero spin without hybridization. For the single phonon mode, $\bm k'=\bm k$ and the band index $\lambda'=\lambda$, the phonon spin reads:
\begin{eqnarray}
\bm S_{\bm k}&=& \langle\bm\varepsilon_{\bm k\lambda}|\hat{\bm{S}}|\bm\varepsilon_{\bm k\lambda}\rangle (b^{\dag}_{\bm k\lambda}b_{\bm k\lambda}+\frac{1}{2}) \nonumber \\
&=&\hbar\text{Im}[\bm\varepsilon^*_{\bm k\lambda}\times\bm\varepsilon_{\bm k\lambda}] (b^{\dag}_{\bm k\lambda}b_{\bm k\lambda}+\frac{1}{2}).
\end{eqnarray}
For the time-reversal phononic (elastic) system, it is readily to observe that:
\begin{equation}\label{symmS}
\bm S_{\bm k}=-\bm S_{-\bm k},
\end{equation}
upon the symmetric phonon excitation $b^{\dag}_{-\bm k\lambda}b_{-\bm k\lambda}=b^{\dag}_{\bm k\lambda}b_{\bm k\lambda}$. This clearly demonstrates the phonon 
spin-momentum locking that reversing $\bm k$ will reverse spin. The valley phonon spin in structures of  $C_3$ symmetry falls into this scenario and opposite valleys have opposite phonon spins upon the inversion $(\bm k\leftrightarrow -\bm k)$, like for the circular polarized phonon modes in hexagonal lattice~\cite{CPphonon1,CPphonon2}. In general, bulk modes of phonons with finite $\bm k$ will have the finite $\bm S_{\bm k}$, manifested as the elliptical polarization. Besides bulk modes, the phonon spins of the topological interface modes, inhomogeneous evanescent waves, and surface acoustic waves, such as Rayleigh-Lamb waves~\cite{long2018intrinsic, yuan2021observation}, all fall into this scenario, and the sign of SAM is directionality dependent. 

So far, we have focused on the phonon spin of particular modes. Now, let us look into the whole phonon spin after the volume integral. By utilizing $\int d\bm r^3 e^{i(\bm k-\bm k')\bm r}=V\delta_{\bm k\bm k'}$, we can obtain
\begin{eqnarray}
\bm S_V&=&\int d\bm r^3 \frac{\rho\omega}{2} \text{Im}[\bm{u}^*\times\bm{u}]     \nonumber \\
&=&\hbar\sum_{\lambda\lambda'}\sum_{\bm k}\text{Im}[\bm\varepsilon^*_{\bm k\lambda}\times\bm\varepsilon_{\bm k\lambda'}](b^{\dag}_{\bm k\lambda}b_{\bm k\lambda}+\frac{1}{2}\delta_{\lambda\lambda'})    \nonumber \\
&=&\sum_{\lambda\lambda'}\sum_{\bm k} \langle\bm\varepsilon_{\bm k\lambda}|\hat{\bm{S}}|\bm\varepsilon_{\bm k\lambda'}\rangle b^{\dag}_{\bm k\lambda}b_{\bm k\lambda'},
\end{eqnarray}
where the zero point phonon spin $\frac{1}{2}\hbar$ is cancelled by the time-reversal pair $\bm k$ and $-\bm k$ due to the symmetry $\langle\bm\varepsilon_{\bm k\lambda}|\hat{\bm{S}}|\bm\varepsilon_{\bm k\lambda'}\rangle=-\langle\bm\varepsilon_{-\bm k\lambda}|\hat{\bm{S}}|\bm\varepsilon_{-\bm k\lambda'}\rangle$.

$\lambda$ and $\lambda'$ can be arbitrary pair of phonon modes, either of the same branch, like previously discussed surface acoustic wave and valley modes, or of different branches, like previously discussed  transverse-transverse hybridization (the simplest example of circularly polarized phonons), longitudinal-longitudinal hybridization, longitudinal-transverse hybridization and degenerate mode hybridization. Therefore, above discussions are also similarly applicable to the volume integrated phonon spin. 
Moreover, because the summation on all $\bm k$, the integrated phonon spin of some cases may vanish due to cancelation between different mode pairs, such as the time-reversal pair, the mirror-reflection pair, the inversion pair, and so on. Therefore, breaking those symmetries or selectively exciting the modes or integrating partial volume is required to obtain nonzero  phonon spin in total. We here confined our discussions about the hybridization and symmetry for phonon spin in a general form. Concrete examples of phonon spin will be demonstrated in a subsequent article for various topological phononic crystals and lattices. %From Elastic Spin to Phonon Spin: Topological Material and Lattice Examples

\section{Conclusion and Discussion}\label{SecVI}
In summary,  by applying the field theory and Noether's theorem to the elastic Lagrangian, we have revisited the elastic spin and orbital angular momentum, and discussed their invariance and symmetry properties. Connections between elastic spin and the vorticity of energy flux and momentum have been discussed. By applying the second quantization to elastic fields, we have discussed general quantum phonon spin. The phonon spin picture uncovered here is not restricted to the transverse phonon modes, but is shown to apply to general phonon modes, like the longitudinal phonon modes, surface phonon modes and hybridized phonon modes, regarded as consequences of mode interferences, whose symmetries and properties are discussed in details. 

%1????????????SOC???types?????????????
%2???T????????????momentum?energy flux??????S???????????????
%3?OAM?SAM????????SAM?????OAM???????????????
%4?????????????????.

Yet, there are many open questions that here we have no space to discuss in detail but deserve further efforts: 

I) From Sec.~\ref{SecI}, it is readily to obtain $\frac{\partial\bm L}{\partial t}+\nabla\cdot(\bm r\times \bm T)=\bm T^T-\bm T$ so that when the stress tensor $\bm T$ is not symmetric as in many cases, the elastic OAM $L_i$ can not be conserved individually. Neither do the elastic SAM $S_i$. Nevertheless, the total AM $\int d\bm r^3 J_{i0}=\int d\bm r^3 (L_i+S_i)$ can still be well conserved in total (see also the $SO(3)$ symmetry discussion in Sec.~\ref{SecII}). This is because considering the invariance $\partial_{\mu} J_{i\mu}=0$ in Eq.~\eqref{totalAM}, the integral $\int d\bm r^3\partial_{0} J_{i0}=-\int d\bm r^3\partial_{k} J_{ik}$ is in the form of a volume integral of a divergence, which is therefore equal to an integral over the bounding surface and vanishes at boundary. The non-conservation of separate OAM and SAM indicates that the elastic field (even general wave system) possesses the spin-orbital coupling (SOC) intrinsically. Our further investigation suggests two different types of SOC: Rashba SOC for the longitudinal wave ($\nabla\cdot \sim \hat{\bm e}_i\cdot(\hat {\bm S}\times \hat {\bm p})_i$) and Dresselhaus SOC for the transverse wave ($\nabla\times \sim \hat {\bm S}\cdot \hat {\bm p}$), which will be discussed in details in future. 

II) The energy-momentum tensor $T^{\nu\mu}$, satisfying the conservation $\partial_\mu T^{\nu\mu}=0$, is indeterminate by any additional term whose spacetime divergence vanishes. When the energy-momentum tensor is asymmetric, people can often exploit the ambiguity to find such a divergenceless quantity to define a symmetrized tensor. As a consequence, the symmetrization of energy-momentum tensor will cause a modification of the definition of momentum and energy flux, as well as the SAM and OAM. Their relations discussed in Sec.~\ref{SecIII} will then be affected. For instance in Eq.~\eqref{PoyntingRelation}, we may have a redefined energy flux $\bm P'=\bm P+v^2_T\nabla\times \bm S$ so that  both Eqs.~\eqref{PT}~\eqref{PL}  may be modified to more symmetric and consistent forms: $\bm P'_T=v^2_T(\bm p^l+\bm p^s)$ and $\bm P'_L=v^2_L(\bm p^l+\bm p^s)$ so that the term $\nabla^2\bm S$ can disappear (be absorbed) in the vorticity expressions of momentum and energy flux. The effects of symmetrization of energy-momentum tensor definitely deserve further studies in future.

III) Hierarchical structure of AM at different scales and perspectives will also cause the possible transformation between SAM and OAM. The infinitesimal element of the continuous medium in the field theory must not be identically recognized as the elementary particle of the corpuscular theories. For particles at location $\bm r+\bm u_i$ with velocity $\dot{\bm r}+\dot{\bm u}_i$, let us denote $\bm r$ being the vector from the origin to the center of mass, $M=\sum_i m_i$ the total mass, and $\bm u_i, m_i$ the displacement vector to the mass center and the mass of the particle $i$. It is well known that (see e.g., Goldstein's textbook about classical mechanics~\cite{Goldstein}) the total AM can be decomposed as $\bm J=\sum_i(\bm r+\bm u_i)\times m_i  (\dot{\bm r}+\dot{\bm u_i})=\bm r\times M  \dot{\bm r}+\sum_i\bm u_i\times m_i  \dot{\bm u}_i$, i.e., the ``external'' AM of motion concentrated at the center of mass, plus the ``internal'' AM around the center of mass. When one can not distinguish the internal structure of the mass, the ``internal'' AM is regarded as the intrinsic spin of the mass. If one has better and better resolving ability, the ``internal'' AM of the mass will be hierarchically found as the resultant of the ``external'' AM of constituent elements, such as molecules, atoms, nuclei, and more deep internal structures. Therefore, the hierarchical structure of substances at different scales causes the transformation of external and internal AM into one another. 

Phonon, as the elementary excitation of the elastic field by second quantization, is a kind of quasiparticle dual to the vibrational wave. Similar to the electromagnetic wave-photon duality, the intrinsic properties of the phonon local in the reciprocal space is related to the extended (global) properties of the wave extended in the real space. A `single' phonon mode is manifested as a collective motion pattern of the whole lattice so that all the properties of a single phonon mode will be flatten throughout the whole real space. Therefore, the intrinsic phonon spin is closely related to the elastic spin, and the density of the latter at different locations is capable of uncovering more rich internal angular momentum structures extended in the real space. 
Investigation of the phonon spin and elastic spin may not only inspire people to pursue the origin of the spin of real fundamental particles, but may also provide us more opportunities to develop {\it spin phononics} for multiple degree of freedom quantum sensing and quantum control technology. When solid-based quantum chips become more and more important in the future, the electron-phonon coupling, opto-mechanical coupling, magneto-elastic coupling, piezoelectric coupling etc. will be inevitably involved due to the substrate vibrations, where the phonon spin and elastic spin will play a central role when coupled to the electron spin, optical spin, magnetic spin, spin wave and so on.  

\begin{acknowledgments}
This work was supported by the National Natural Science Foundation of China (Grant No. 11935010), the National Key R\&D Program of China (2022YFA1404400) and the Opening Project of Shanghai Key Laboratory of Special Artificial Microstructure Materials and Technology.
\end{acknowledgments}

\bibliography{references}
\end{document}